\documentclass[12pt]{article} 
\usepackage{amsmath,amsfonts, amssymb}
\usepackage{comment}
\usepackage{graphicx}
\usepackage{amssymb}
\usepackage{epstopdf}
\newcommand{\be}{\begin{equation}}
\newcommand{\ee}{\end{equation}}
\newcommand{\ba}{\begin{eqnarray}}
\newcommand{\ea}{\end{eqnarray}}

\begin{document}

\vskip   1cm
\begin{center} {\Large{\textsc{Study of the OSV Conjecture for 4D N=2 Extremal Black Holes in Type IIB String Theory} }}
\end{center}
\vskip 1cm
\renewcommand{\thefootnote}{\fnsymbol{footnote}}
\centerline{\bf Farzad Omidi \footnote{omidi@alum.sharif.edu}}
\vskip 2cm
\centerline{\small Department of Physics, Sharif University of Technology,}
\centerline{ \small  P.O. Box 11155-9161, Tehran- Iran }
\bigskip

\begin{abstract}
In this survey we study the OSV conjecture for 4D N=2 extremal black holes of type IIB suprstring theory. We apply T-duality to find the generalized prepotential of the low energy limit of this superstring theory up to one-loop order in the closed string coupling. On the other hand, we calculate the tree-level and one-loop free energies of B-model topological string theory. 
To compare them we will explicitly show that the OSV conjecture holds for type IIB N=2 extremal black holes.
\end{abstract}
 
\newpage
\section{Introduction}
Topological Sigma models and string theories were introduced and mainly developed by Edward Witten by the end of 1991 ~\cite{witten1,witten2,vafamir,vafarev}. Although the main motivation for their introduction was to find a way to describe topological invariants of certain manifolds by correlation functions of some kinds of field theories; gradually more physicists became interested in studying and applying them. The first physical application of topological string theories were found in 1994 by Antoniadis, Gava, Narain, and Taylor ~\cite{antoni}. They could successfully show that the g-loop scattering amplitude of 2 gravitons and 2g-2 graviphotons in type IIA (IIB) superstring theory comapctified on a Calabi-Yau Manifold $M$ is equal to the g-loop partition function of A (B) model topological string theory which lives on the same Calabi-Yau manifold $M$.
\begin{equation}
\langle {R^2 T^{2g-2}}\rangle ^{IIA/IIB}_{g-loop}=F_g^{topA/topB}(t^{\alpha})
\end{equation}
in this expression R  and T are the field strengths of graviton and graviphoton respectively; $F_g^{top}(t)$ is the g-loop topological string partition function. For A (B) model topological string theory $t^\alpha$ are k\"ahler (complex) structure moduli of the Calabi-Yau which is used in the compactification of the corresponding superstring theory. Moreover, Bershadsky, Cecotti, Ooguri, and Vafa could prove that
\begin{equation}
{{\langle} {R^2}{\rangle}}_{one-loop}^{IIA}=F_{g=1}^{top A}(t^{\alpha})
\end{equation}  
On the other hand, Cardoso, de Wit, and Mohaupt applied Wald's formula  and computed the entropy of these black holes ~\cite{dewit1,devia,moha}. According to (1) the g-loop free energies of topological string theories clearly appear in the low energy effective action of type II supersting theories. In other words, one has terms such as 
\begin{equation}
S_{effective}^{IIA/IIB}=\sum_{g=1}{g{F_g}^{top A/top B}}[R^2 T^{2g-2}+2(g-1)(RT)^2 T^{2g-2}]
\end{equation}
where $R^2$ means $R_{\mu\nu\rho\sigma}R^{\mu\nu\rho\sigma}$, $(RT)^2=(RT)_{\rho\sigma}(RT)^{\rho\sigma}$, and $(RT)_{\rho\sigma}={R_{\mu\nu\rho\sigma}T^{\mu\nu}}$.
Therefore, it looks reasonable that one can express the entropy or partition function of the BPS black hole in terms of an observable in the corresponding topological string theory. Finally, in 2004 Ooguri, Strominger, and Vafa conjectured one can define a mixed ensemble for 4D N=2 BPS black holes in type II superstring theories, and write a partition function for them ~\cite{osv}. Theses black holes are described by three observables comprising their mass $M$, electric charges $q_I$, magnetic charges $p^I$. More precisely, they considered electric and magnetic charges as two different degrees of freedom and utilized a microcanonical ensemble for magnetic charges and a canonical ensemble for electric charges. In this manner, one can write the partition function of the black hole, and compute its free energy.
\begin{eqnarray}
Z_{mixed}= \sum_{q _{I}} \Omega(p^I, q _{I})\times \exp (-q.\phi)
\end{eqnarray}
in which $\Omega(p^I, q_{I})$ indicates the degeneracy of states with magnetic charges $p^I$ and electric charges $q_I$, and $\phi^I$ specify how much energy of the system will be changed if one adds or subtracts a unit of electric charge to the black hole. Since, the meaning of ${\phi}^I$ is analogous to that of chemical potential in statistical physics, Ooguri, Strominger, and Vafa (OSV) called it the electric chemical potential. Similarly, for every magnetic charge there is a magnetic chemical potential which in the literature is shown by $\chi_I$.
\\ When one compactifies type IIA or IIB supersting theory on a Calabi-Yau manifold, whose holonomy group is $SU(3)$, obtains some massless scalar fields $X^I$ from components of different tensor fields. Some of them sit in N=2 vector multiplets, others sit in N=2 hypermultiplets. Theses massless scalar fields called moduli are k\"ahler or complex structure moduli of the Calabi- Yau manifold.
Here we show K\"ahler moduli by $X^I$, and complex structure moduli by $\acute{X^I}$. The space of Ka\"eler moduli itself has proven to be a k\"ahler manifold; this manifold has a real nowhere vanishing function named K\"ahler potential $K(X)$ whose second order derivatives give the components of the K\"ahler metric, $g_{i\bar{j}}=\partial_i\partial_{\bar{j}} K$ of the K\"ahler moduli space. 
\\There is a function (or more precisely a section) $F(X^I)$ of moduli whose first order derivatives give the K\"ahler potential $K(X)$. It is called prepotential since its first derivatives determine the K\"ahler potential ~\cite {moha}.
\begin{equation}
K(X)=-\ln (-i[X^I\bar{F_I}-F_I \bar{X^I}])
\end{equation}
It is useful to introduce the inhomogeneous coordinates $t^{\alpha\neq0}=\dfrac{X^{I\neq0}}{X^0}$ on the moduli spaces.
The prepotential receives loop corrections in closed string coupling; after taking into account theses loop corrections, it is called generalizaed prepotential. For type IIA supergravity and up to the one loop order in string coupling it is given by ~\cite{devia}
\begin{equation}
F^{IIA}(t^\alpha)=(CX^0)^2 D_{\alpha\beta\gamma}t^{\alpha}t^{\beta}t^{\gamma}-\dfrac{1}{6}c_{2\alpha}t^{\alpha}+....
\end{equation}
the first term is computed at tree level in string coupling and $ D_{\alpha\beta\gamma}$ determines the number of intersections of four cycles of the Calabi-Yau manifold. If one chooses bases $ \lbrace\omega_{\alpha}\rbrace $ for the cohomology group $H^{(1,1)}(M)$ it can be written as
\begin{equation}
D_{\alpha\beta\gamma}=\int_M \omega_{\alpha}\wedge\omega_{\beta}\wedge\omega_{\gamma}
\end{equation}
The second term in (6) occurs at one-loop order in string coupling, and is related to the second Chern class $ c_2 $ of the Calabi-Yau manifold.
\begin{equation}
c_{2\alpha}=\int_M c_2 \wedge \omega_\alpha
\end{equation}
On the other hand, the free energy of A model topological string theory has a perturbative expansion in closed topological string coupling $ g_{top} $ as follows [7,11]
\begin{equation}
F^{top,A}(t^\alpha)=-\dfrac{{(2\pi)}^3 i}{g_{top,A}^2} D_{\alpha\beta\gamma}t^{\alpha}t^{\beta}t^{\gamma}-\dfrac{\pi i}{12}c_{2\alpha}t^{\alpha}+...
\end{equation}
By comparing (6) and (9) term by term, OSV concluded that 
\begin{equation}
F^{IIA}(t^{\alpha})=-\dfrac{2i}{\pi}F^{top,A}(t^\alpha)
\end{equation}
Moreover, by utilizing the mixed ensemble, OSV could prove that the free energy of N=2 extremal black holes is obtained by taking the imaginary part of the generalized prepotential. 
Therefore, they gained the following astonishing  formula which relates the partition function $ Z^{IIA}_{Mixed} $ of 4D N=2 extremal Type IIA black holes to the partition function $ Z^{top,A} $ of the A-model topological string theory. 
\begin{equation}
Z^{IIA}_{Mixed} =\vert{ Z^{top,A} }\vert^2
\end{equation}
Here we should emphasize on few points. First, the corresponding topological string theory lives on the same Calabi-Yau manifold on which Type IIA superstring theory is compactified. Second, two-loop and higher order corrections in the generalized prepotential and in the topological string theory have been ignored. It can be shown that by choosing a special charge configuration, in which some of electric and magnetic charges are very large, such higher order corrections can be omitted ~\cite{osv}. Third, though in ~\cite{osv} authors used a mixed ensemble to derive equation (11), shortly after that, people could prove that one can use any kind of ensemble to describe the black hole ~\cite{ensemble}. This situation is analogous to what we have in ordinary statistical mechanics; All ensembles are equivalent to each other in an appropriate limit. In other words, no matter of which ensemble one chooses to compute the thermodynamical quantities of the system, the results derived by different ensembles should be equal. In this context, the appropriate limit is the limit of large electric and magnetic charges, i.e. in this limit the microcanonical and canonical ensembles become equivalent to each other.
\\The importance of this discovery is that calculations in topological string theories are much easier than in ordinary superstring theories. The reasons are as follows: First, Topological string theories can merely live on Euclidean, i.e. Calabi-Yau, manifolds. Hence they are time independent. Second, Their Hilbert spaces are much smaller. Third, their vertex operators are in a one to one correspondence to the members of the Dolbeualt cohomology groups $ H^{(1,1)}(M) $ or $ H^{(2,1)}(M) $ in A-model or B-model topological string theories, respectively. Consequently, computing correlation functions are reduced to the calculation of the integrals of the Wedge product of theses forms on the Calabi-Yau manifold. Due to these facts, this conjecture has excited a lot of interest during the past six years.
\\ Having been reviewed the original conjecture, we are now ready to study this intriguing idea in the case of 4D N=2 extremal blcak holes in the low energy limit of type IIB superstring theory. 
\newpage
We know if one compatifies type IIA superstring theory on a circle of radius $R$, and applies T-duality on the circle, then type IIA theory turns into type IIB theory on a circle whose radius is $\dfrac{1}{R}$. In 1994, Zaslow and Strominger ~\cite{zaslow} proved that T-duality is a special case of a symmetry named Mirror Symmetry. Mirror symmetry turns a Calabi-Yau threefold  $M$ into another Calabi-Yau threefold $\acute{M}$ (called mirror manifold) such that there is an isomorphisms between certain Dolbeault cohomology groups ~\cite{vafamir,mir1},
\begin{equation}
 H^{(1,1)}(M)  \cong  H^{(2,1)}(\acute{M})
\end{equation}
and their Euler characteristics are related to each other by:
\begin{equation}
\chi =- \acute{\chi} 
\end{equation}

These formulae just hold for the case in which the complex dimension of $M$ is three. Under Mirror symmetry type IIA superstring theory compactified on $M$ is exchanged by type IIB superstring theory compactified on $\acute{M}$. 
\\On the other hand, since mirror symmetry exchanges K\"ahler structure moduli $t^{\alpha}$ of manifold $M$ and complex structures moduli $\acute{t}^{\alpha}$ of manifold $\acute{M}$, through the isomorphism (12), it turns the type A topological string theory on $M$ into type B topological theory on $\acute{M}$ ~\cite{vafamir, vafarev}. By putting everything together we can draw figure 1. Hence it seems probable that one can write the partition function of 4D N=2 extremal black holes of type IIB in terms of the partition function of B-model topological string theory. Our aim in this paper is to verify this claim explicitly. 

\begin{figure}
\centering
\includegraphics[scale=.6]{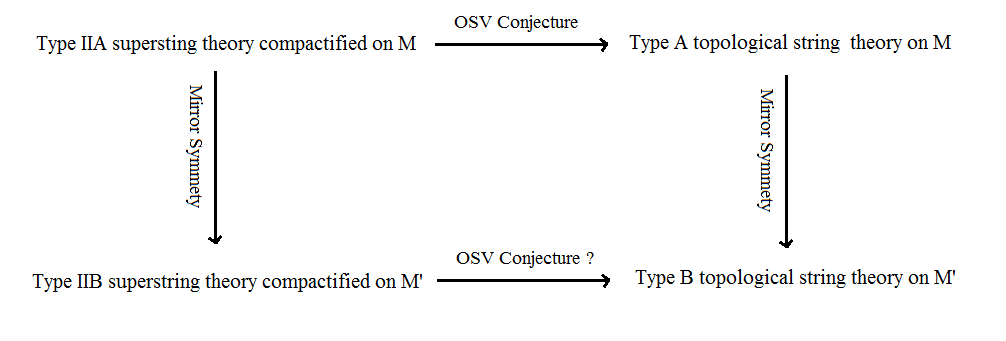}
\caption{Mirror Symmetry supports the presence of the OSV cojecture for N=2 extremal black holes of type IIB}
\label{fig:1}
\end{figure}
The paper is organized as follows: In section 2 we derive the prepotential of type IIB supergravity up to one loop order in string coupling. In section 3 we calculate the free energy of type B topological string theory up to one loop order. We discuss that one needs a field redefinition in order to calculate the one loop free energy. In section 4 we compare the prepotential and the free energy that we obtained in the previous sections. At the end by utilizing a mixed ensemble for the black hole we verify that the OSV conjecture holds for 4D N=2 extremal black holes of type IIB string theory. To start we first need to know the generalized prepotential of the low energy limit of 4D type IIB string theory.

\section{Perepotential of Type IIB Supegravity}

 To find the prepotential we will use mirror symmetry. According to (12) we have
 \begin{equation}
 \omega \in H^{(1,1)}(M)\longrightarrow \xi \in H^{(2,1)}(\acute{M})
 \end{equation}
 On the other hand, there is another isomorphism 
 \begin{equation}
 H^{(2,1)}(\acute{M})\cong H^{(0,1)}(\acute{M},T\acute{M})
 \end{equation}
A member $\acute{\mu}$  of $ H^{(0,1)}(\acute{M},T \acute{M}) $ is a closed anti-holomorphic 1-form as well as a holomorphic vector, i.e. $\acute{\mu}=\acute{\mu}^l_{\bar{k}} d \acute{\bar{z}}^{\bar{k}} \acute{\partial_l}$. Consequently, isomorphism (12) takes the K\"ahler form $\omega \in H^{(1,1)}(M)$ (13) to $\acute{\mu} \in H^{(0,1)}(\acute{M},T\acute{M})$ :
 \begin{equation}
 \omega \longrightarrow \acute{\mu}  
 \end{equation}
One can write both sides of the above isomorphism in terms of the the components of the forms, 
\begin{equation}
 \omega=iG_{k\bar{k}}dz^k \wedge dz^{\bar{k}} \longrightarrow \acute{\mu}= \acute{G}_{k\bar{k}} \acute{\mu}^k _{\bar{l}} {\acute{G}^{l\bar{l}} d \acute{\bar{z}}^{\bar{k}} \acute{\partial_l}}
\end{equation}
in which $ \acute{z} $ and $ \acute{G} $ are the coordinates and the K\"ahler metric of the mirror manifold $ \acute{M} $, respectively. On the other hand, it has been proven that under this symmetry the K\"ahler metric and anti-symmetric tensor of the manifold $M$ transform as follows ~\cite{Giveon, mir2}:
\begin{eqnarray}
G &\longrightarrow& \acute{G}(G,B)
\\ \nonumber B &\longrightarrow& \acute{B}(G,B)
\end{eqnarray}
In other words, the new metric $\acute{G}$ is a function of both $G$ and $B$. By using (17) and (18) one can write:
\begin{equation}
dz^k \wedge d{\bar{z}}^{\bar{k}}\longrightarrow -i \acute{\mu}^k_{\bar{l}} \acute{G}^{l \bar{l}} d{\acute{\bar{z}}}^{\bar{k}} \acute{{\partial}}_{\bar{l}}
\end{equation}
Thus the Ricci form $R_{j}^{i} $ and the second Chern class $c_{2} $ of the manifold $M$ transform in the following way 
\begin{equation}
R_{j}^{i}\longrightarrow \acute{R}_{j}^{i}:=-i \acute{R}^i_{jk\bar{k}}  \acute{\mu}^k_{\bar{l}} \acute{G}^{l \bar{l}} d{\acute{\bar{z}}}^{\bar{k}} \acute{{\partial}}_{\bar{l}}
\end{equation}
\begin{equation}
c_2 \longrightarrow \acute{c_2}:=\dfrac{1}{8{\pi}^2} {\delta}_{i_1i_2}^{j_1j_2}  \acute{R}_{j_1}^{i_1} \wedge \acute{R}_{j_2}^{i_2} 
\end{equation}
Putting all these together we have
 \begin{equation}
 D_{\alpha\beta\gamma} \longrightarrow \acute{D}_{\alpha\beta\gamma}:= \int_{\acute{M}} \acute{\mu}_{\alpha} \wedge \acute{\mu}_{\beta} \wedge \acute{\mu}_{\gamma} \, \ \acute{\Omega}_{ijk} \wedge \acute{\Omega}
\end{equation}
and 
\begin{equation}
c_{2\alpha} \longrightarrow \acute{c_{2\alpha}}:=\int_{\acute{M}} \acute{c}_2 \wedge {\acute{\mu}}_{\alpha}
\end{equation}
Here $ {\acute{\mu}}_{\alpha} $  are bases for the cohomology group $H^{(0,1)}(\acute{M},T \acute{M})$ and called Beltrami differentials. Note also that $  \int_{\acute{M}} {\acute{\mu}}_{\alpha} \wedge {\acute{\mu}}_{\beta} $ 
\\Generally the map which relates the K\"ahler and complex structure moduli of both Calabi-Yau manifolds is very complicated ~\cite{mir1}. But if one considers the case in which the volume of $M$ and $\acute{M}$ are both large, then this map is linear. 
\begin{equation}
\acute{t}^{\alpha}=\sum_{\beta}L_{\alpha\beta}t^{\beta}
\end{equation}
in which the coefficients $L_{\alpha\beta}$ are independent of the moduli. 
Finally, we will find 
\begin{equation}
F^{IIB}(\acute{t}^{\alpha}) ={(\acute{C}\acute{X}^0)}^2 \acute{D}_{\alpha\beta\gamma}\acute{t}^{\alpha}\acute{t}^{\beta}\acute{t}^{\gamma}-\dfrac{1}{6}\acute{c}_{2\alpha} \acute{t}^{\alpha}+...
\end{equation}
\section{Free Energy of Type B Topological String Theory}
Now we want to examine the free energy of type B topological string theory. The tree level term has already been derived ~\cite{vafamir}. According to ~\cite{kod,vonk} the tree level free energy is zero and one has to insert three vertex operators in the path integral to get a non-zero result. In other words, at tree level in topological string coupling the first non-vanishing observable is the three point function which is ~\cite{vafamir}:
\begin{equation}
F^{top,B}_{tree-level} (\acute{t}^\alpha)= \int_{\acute{M}} (\acute{\mu})^i \wedge (\acute{\mu})^j (\acute{\mu})^k {\Omega}_{ijk} \wedge \Omega
\end{equation}
Here $(\acute{\mu})^i$ belongs to $H^{(0,1)}(\acute {M},T\acute{M})$. By choosing bases $\lbrace{(\acute{\mu}_\alpha)^i}\rbrace $ for $H^{(0,1)}(\acute {M},T\acute{M})$ and expanding  $(\acute{\mu})^i$ in terms of these bases,
\begin{equation}
(\acute{\mu})^i=(\acute{\mu}_{\alpha})^i\acute{t}^\alpha,
\end{equation}
we can rewrite (27) as
\begin{equation}
F_{tree-level}^{top,B}(\acute{t}^\alpha)=\acute{D}_{\alpha\beta\gamma} \acute{t}^{\alpha} \acute{t}^{\beta} \acute{t}^{\gamma}
\end{equation}
where $ \acute{D}_{\alpha\beta\gamma} $ is defined by (23). The next step is to calculate the one-loop free energy $ F_{one-loop}^{top,B} $ of the topological string theory. To do so we should calculate the following index ~\cite{one}
\begin{equation}
F_{one-loop}^{top,B}=\dfrac{1}{2} \int_F  \dfrac{d^2\tau}{{\tau}_2} Tr[(-1)^F F_L F_R q^{H_L} {\bar{q}}^{H_R}]
\end{equation}
where the integral is taken on the fundamental domain of the worldsheet which is a torus. $ F_L $ and $F_R$ are charges corresponding to $U(1)$ R symmetry in the left-moving and right-moving N=(2,2) supersymmetry algebra. $H_L$ and $H_R$ are the left-moving and right-moving parts of the Hamiltonian of B-model topological string theory. To compute it we proceed like ~\cite{one} and turn it into a path integral over permitted field configurations. We know from ~\cite{witten1,vafamir} that as a result of a fixed point theorem only constant bosonic fields have a non-zero contribution in the path integral. Furthermore, we know that observables of this theory are independent of the k\"ahler structure moudli (or the volume) of $\acute{M}$ and depend on complex structure moduli (or the shape) of $\acute{M}$. Hence, we can work in the limit where $\acute{M}$ is very large. The  Lagrangian density is as follows ~\cite{witten1} 
\begin{eqnarray}
L_{top,B}&=&G_{IJ} \partial_I {\phi}^I \partial_J {\phi}^J+iG_{i\bar{j}} {\eta}^{\bar{j}}(D_{\bar{z}}{\rho}_z^i+D_z{\rho}_{\bar{z}}^i) 
\\ \nonumber &+& i{\theta}_i(D_{\bar{z}}{\rho}_z^i-D_z{\rho}_{\bar{z}}^i)+R_{i\bar{j} k}^l {\rho}_z^i {\rho}_{\bar{z}}^k {\eta}^{\bar{j}} {\theta}_l 
\end{eqnarray}
 Here $G_{i \bar{j}}$ and $R_{i\bar{j} k}^l $ are the metric and the Riemann tensor of the Calabi-Yau three-fold $\acute{M}$. ${\phi}^{I=i,\bar{i}}$ are bosonic fields on the topological string world-sheet. Moreover, ${\rho}_z^i$ and ${\rho}_{\bar{z}}^i$ are the anti-commuting spin 1 and -1, $\theta_i$ and $\eta^{\bar{j}}$ the anti-commuting spin 0 fields on the world-sheet. $D_z$ and $D_{\bar{z}}$ are covariant derivatives:
 \begin{equation}
 D_z \rho_{\bar{z}}^i=\partial_z \rho_{\bar{z}}^i+\Gamma_{jk}^i \partial_z \phi^j \rho_{\bar{z}}^k
 \end{equation}
By looking at the Lagrangian density (31), We see that in order to prevent the action from being infinite and hence getting a zero partition function, all the bosonic and fermionic fields must be constant. Therefore, one should only consider the four-ferminon interaction term in the action.
 \\By applying Noether's theorem, one can easily obtain R-symmetry charges 
 \begin{eqnarray}
 F_L&=& -2\pi (G_{i\bar{i}} {\eta}^{\bar{i}}+{\theta}_i) {\rho}^i_{\bar{z}}
 \\ \nonumber F_R &=& -2\pi (G_{i\bar{i}} {\eta}^{\bar{i}}-{\theta}_i) {\rho}^i_z
 \end{eqnarray} 
 In analogy with ~\cite{one} we have 
 \begin{eqnarray}
& F_{one-loop}^{top,B} =  \int_{F_0}  \dfrac{d^2\tau}{4 \pi{{\tau}_2}^4} \int_{\acute{M}} dV \int {\prod}_r d{\rho}^r_z d{\rho}^r_{\bar{z}}
 d{\eta}^{\bar{r}} d{\theta}_{r}&
 \\ \nonumber &\times {\rho}_z^k {\rho}_{\bar{z}}^l (G_{l\bar{l}} {\eta}^{\bar{l}}+{\theta}_l)(G_{k\bar{k}} {\eta}^{\bar{k}}-{\theta}_k)
 \exp(\pi {\tau}_2 R_{i\bar{i} j \bar{j}} {\rho}^i_{\bar{z}} {\rho}^j_z {\eta}^{\bar{i}} {\theta}_k G^{k\bar{j}})&
 \end{eqnarray}
 In this expression $dV$ is the volume element of $\acute{M}$. By using Taylor's expansion and keeping in mind that we have a Berezin integration, only the second term in this expansion will have a non-zero contribution in the path integral. 
 \begin{equation}
 \dfrac{1}{32\pi} \int_{F_0}  \dfrac{d^2{\tau}}{({\tau}_2)^2} \int_{\acute{M}} dV {\epsilon}^{l i_2 i_3} {\epsilon}^{k j_2 j_3} {\epsilon}^{\bar{l} \bar{i}_2 \bar{i}_3} {\epsilon}^ {\bar{n} \bar{j}_2 \bar{j}_3} G_{l\bar{l}} \,G_{k \bar{n}} R_{i_2 \bar{i}_2 j_2 \bar{j}_2} R_{i_3 \bar{i}_3 j_3 \bar{j}_3}
 \end{equation}
 Taking the integral over the fundamental domain $F_0$ of the genus one worldsheet, and using the following relations
 \begin{equation}
 {\epsilon}^{i_1 i_2 i_3}{\epsilon}_{i_1 j_2 j_3}={\delta}_{j_2 j_3}^{i_2 i_3}
 \end{equation}
 \begin{equation}
\acute{c}_2=-\dfrac{1}{8{\pi}^2}{\delta}_{i_1 j_2}^{j_1 j_2} \acute{R}_{j_1}^{i_1} \wedge \acute{R}_{j_2}^{i_2}
 \end{equation}
\begin{equation}
\int_{\acute{M}}  \acute{\omega} \wedge \acute{c}_2 =\dfrac{1}{8{\pi}^2} \int dV {\epsilon}^{i_1 i_2 i_3} {\epsilon}^{j_1 j_2 j_3} {\epsilon}^{\bar{i}_1 \bar{i}_2 \bar{i}_3} {\epsilon}^ {\bar{j}_1 \bar{j}_2 \bar{j}_3} \acute{G}_{i_1\bar{i}_1} \, \acute{G}_{j \bar{j}_1} \acute{R}_{i_2 \bar{i}_2 j_2 \bar{j}_2} \acute{R}_{i_3 \bar{i}_3 j_3 \bar{j}_3} 
 \end{equation} 
 We can write
 \begin{equation}
 F_{one-loop}^{top,B}=\dfrac{\pi}{4}\int_{\acute{M}} \acute{\omega} \wedge \acute{c}_2
 \end{equation}
 It is obvious that this expression linearly depends on the K\"ahler form $ \acute{\omega}$, and hence on the K\"ahler structure moduli of $\acute{M}$. Since it must depend on the complex structure moduli of $\acute{M}$, we certainly made a mistake! To understand what has led us to this wrong result, we should look at (35). In this expression there are two metrics: one is absorbed in the contraction of two Levi-Civita symbols (or more precisely in the definition of $\acute{c}_2$), and the other one is used in the K\"ahler form $\acute{\omega}$. In other words, the only dependence of (40) on k\"ahler moduli comes from the K\"ahler form $\acute{\omega}$ or equivalently from the metric $ g_{l\bar{l}} $.
 \\To reach the right result, we have to do something to replace $\acute{\omega}$ with another form which depends on the complex structure moduli $\acute{t}^\alpha$ of $\acute{M}$. We know that for compact Calabi-Yau manifolds the components of holomorphic three-form $\acute{\Omega}$ and its complex conjugate are constant ~\cite{vafamir}. Furthermore, (1,1)-forms and (2,2)-forms, which are related by Poincar$\acute{e}$ duality, depends on the K\"ahler moduli. Thus, just the members of cohomology group $H^{(2,1)}(\acute{M})\cong H^{(0,1)}(\acute{M}, T\acute{M})$ (or the Belterami differentials $\mu$) depend on the complex structure moduli. Since, $ g_{l\bar{l}} $ has been resulted from the product $F_L F_R$, we should change $F_L$ and $F_R$ such that their product comprises the components of $\mu$. we can accomplish this by field redefinition of one of the Fermionic fields $ \lbrace {\rho}_z^i, {\rho}_{\bar{z}}^i, {\eta}^{\bar{i}}, {\theta}_i \rbrace $, so that in the new definition the components of Belterami differentials $\mu$ are used.
 \\ You may remember that in the first paper ~\cite{witten1} in which Witten introduced type B topological sigma model, he applied the following field redefinition, in order to manifest the new spin of Fermionic fields after twisting, and to simplify the appearance of BRST-like transformations. These transformations act on these fields in the following way:
 \begin{eqnarray}
 \nonumber \delta {\phi}^i&=&0,
 \\ \nonumber \delta {\phi}^{\bar{i}}&=&\epsilon{\eta}^{\bar{i}},
 \\  \delta {\eta}^{\bar{i}}&=&\delta{\theta}_i =0,
 \\ \nonumber \delta {\rho}_z^i &=&2i\epsilon {\partial}_z {\phi}^i,
 \\ \nonumber \delta {\rho}_{\bar{z}}^i &=&-2i\epsilon \bar{{\partial}}_{\bar{z}} {\phi}^i
 \end{eqnarray}One can see that if the field ${\eta}^{\bar{i}}$ is contracted by the components of $\acute{\mu}$,
 \begin{equation}
 {\eta}^{\bar{i}} \longrightarrow {\alpha}^l:=\acute{\mu}_{\bar{i}}^l {\eta}^{\bar{i}},
 \end{equation}
 then the variation of $ {\alpha}_i$, and hence the new vertex operators,
 \begin{equation}
 O_B^{(0)}=B_{i_1...i_p}^{j_1...j_q} {\alpha}^{i_1}...{\alpha}^{i_p} {\theta}_{j_1}....{\theta}_{j_q}
 \end{equation}
 would not be zero under the transformations (40). In (41) $B_{i_1...i_p}^{j_1...j_q}$ are the components of forms which belong to $H^{(p,0)} ({\wedge}^q T^{(1,0)} \acute{M})$. We could do the same redefinition for ${\theta}_i$ or for a linear combinations of them. They all lead to the same incorrect vertex operators, since we know that the correct ones belong to $H^{(0,p)}({\wedge}^q T^{(1,0)}\acute{M})$. Therefore, we can only change the definition of the field ${\rho}_z^i$ or ${\rho}_{\bar{z}}^i$, in order to have the correct vertex operators.
 \begin{equation}
 {\rho}_z^j \longrightarrow {\rho}_z^{\bar{i}}:={\mu}_j^{\bar{i}}  {\rho}_z^j
 \end{equation}
 or
  \begin{equation}
 {\rho}_{\bar{z}}^j \longrightarrow {\rho}_{\bar{z}}^{\bar{i}}:={\mu}_j^{\bar{i}}  {\rho}_{\bar{z}}^j
 \end{equation}
 If one considers the matrix ${\mu}_{\bar{i}}^j$, then ${\mu}_j^{\bar{i}}$ is its inverse matrix. One can show that (43) and (44) will give the desired result. Since, they have no advantages to each other and we cannot use both of them simultaneously, we choose the following redefinitions.
 \begin{eqnarray}
 {\gamma}^i&:=&{\rho}_z^i-{\rho}_{\bar{z}}^i,
\\ \nonumber {\beta}^{\bar{j}}&:=&{\mu}_i^{\bar{j}} ({\rho}_z^i+{\rho}_{\bar{z}}^i),
 \end{eqnarray}
 With this field redefinition charges will change to
 \begin{eqnarray}
 F_L&=&-\pi (G_{i \bar{i}} {\eta}^{\bar{i}}+\theta_i)(\mu_{\bar{j}}^i{\beta}^{\bar{j}}-{\gamma}^i)
 \\ \nonumber  F_R&=&-\pi (G_{i \bar{i}} {\eta}^{\bar{i}}-{\theta}_i)({\mu}_{\bar{j}}^i{\beta}^{\bar{j}}+{\gamma}^i)
 \end{eqnarray}
 Moreover the interaction term in the action changes
 \begin{equation}
 \dfrac{1}{8}R_{i_2\bar{i}_2 j_2\bar{j}_2}\mu_{\bar{k}_2}^{i_2} \beta^{\bar{k}_2} \gamma^{j_2} \eta^{\bar{i}_2} \theta_{k_2} G^{k_2\bar{j}_2}
 \end{equation}
Plugging (46) and (47) into (30) and taking the Grassmannian integrals, the result will be the following expression.
\begin{eqnarray}
&\dfrac{\pi}{64} \int_{F_0}  \dfrac{d^2{\tau}{({\tau}_2)^2}} \int_{\acute{M}} dV \mu_{\bar{i}}^e [\epsilon ^{\bar{i} \bar{i}_2 \bar{i}_3} \epsilon _l^{\bar{j}_2 \bar{j}_3}+\epsilon^{\bar{i} \bar{j}_2 \bar{j}_3} \epsilon _l^{\bar{i}_2 \bar{i}_3}]& 
\\ \nonumber &\times \epsilon_{el_{2}l_{3}} \epsilon^{lj_2j_3}(R_{i_2\bar{i}_2j_2\bar{j}_2} \mu_{\bar{k}_2}^{i_2} G^{\bar{k}_2l_2})(R_{i_3\bar{i}_3j_3\bar{j}_3} \mu_{\bar{k}_3}^{i_3} G^{\bar{k}_3l_3})&
\end{eqnarray}
If we change the indices as
\begin{equation}
\bar{i}_a\longleftrightarrow \bar{j}_a \qquad            a=2,3
\end{equation}
and utilize the equality $R_{i\bar{i}j\bar{j}}=R_{i\bar{j}j\bar{i}}$, we will see that the second term in the brackets will be equal to the first one. Thus, with this field redefinition $ G_{l\bar{l}} $ in (34) is replaced by the components of $\mu$ in (48). Now, we apply the the definitions (21) and (22) for $\acute{c}_2$, and take advantage of the fact that for a compact Calabi-Yau manifold $\Omega_{nl_2l_3}=\epsilon_{nl_2l_3}$ is constant with respect to the coordinates on the manifold. Hence we can rewrite (48) as follows:
\begin{equation}
F_{one-loop}^{top,B}=-\dfrac{{(\pi)}^4}{12}i\acute{t}^\alpha 
\int_{\acute{M}} \Omega_{rr_2r_3}(\mu_\alpha)^r \wedge (\acute{c}_2)^{r_2r_3} \wedge \Omega
\end{equation} 
By using the definition (24) we have
\begin{equation}
F_{one-loop}^{top,B}=-\dfrac{{(\pi)}^4}{12} i\acute{t}^\alpha \acute{c}_{2\alpha}
\end{equation}
On the other hand, one can write the total free energy $ F^{top,B}$ of the type B topological string theory as 
\begin{equation}
F^{top,B}=\Sigma_{g=0}^{\infty} (g_{top,B})^{2g-2} F_g ^{top,B}(\acute{t}^{\alpha})
\end{equation}
Since the coupling $ g_{top,B}$ is very small in this case and the complex structure moduli $\acute{t}_{\alpha}$ are very large, only the first two terms are important. Thus we merely consider the contribution of theses terms. Plugging (29) and (51) into the (52) we obtain the free energy of B-model topological string theory up to one loop order in closed topological string coupling.
\begin{equation}
F^{top,B}=-\dfrac{-6i}{{(g_{top,B})^2}}\acute{D}_{\alpha\beta\gamma} \acute{t}^{\alpha} \acute{t}^{\beta} \acute{t}^{\gamma}-\dfrac{{(\pi)}^4 i}{12} \acute{c}_{2\alpha}  \acute{t}^\alpha +...
\end{equation}
One can change the normalization of the worldsheet fields so that $F^{top,B} $ can be written as 
\begin{equation}
F^{top,B}=-\dfrac{-(2\pi)^3 i}{{(g_{top,B})^2}}\acute{D}_{\alpha\beta\gamma} \acute{t}^{\alpha} \acute{t}^{\beta} \acute{t}^{\gamma}-\dfrac{{\pi i}}{12}  \acute{c}_{2\alpha}  \acute{t}^\alpha+...
\end{equation}

\section{OSV Conjecture for Type IIB N=2 Extremal Black Holes}
In the low energy limit of type IIB superstring theory compactified to four dimensions one can build N=2 extremal black holes in four dimension, by making a bound states of D1, D3, and D5 branes wrapped on cycles of the Calabi-Yau three-fold. Since these branes have mass as well as electric and magnetic charges, the final result will be an extremal black hole in four dimensions.
Again with the same reasoning one can attribute a mixed ensemble to it, and write its free energy as the imaginary part of the generalized prepotential.
\begin{equation}
F_{OSV}^{IIB}=-\pi Im(F^{IIB}(\acute{t}))
\end{equation}
It is very obvious that the type IIB prepotential 
\begin{equation}
F^{IIB}(\acute{t}^{\alpha}) =(\acute{C}\acute{X^0})^2\acute{D}_{\alpha\beta\gamma}\acute{t}^{\alpha}\acute{t}^{\beta}\acute{t}^{\gamma}
-\dfrac{1}{6}\acute{c}_{2\alpha} \acute{t}^{\alpha}+...
\end{equation}
and the free energy of typ topological string
\begin{equation}
F^{top,B}(\acute{t}^{\alpha})=-\dfrac{(2 \pi)^3 i}{{(g_{top,B})^2}}\acute{D}_{\alpha\beta\gamma} \acute{t}^{\alpha} \acute{t}^{\beta} \acute{t}^{\gamma}-\dfrac{\pi i}{12}  \acute{c}_{2\alpha}  \acute{t}^\alpha+...
\end{equation}
are very similar to each other, and must be equal to each other up to a coefficient. Like in ~\cite{osv} we first compare the one loop terms of them and obtain
\begin{equation}
F^{IIB}(\acute{t}^{\alpha})=-\dfrac{2i}{\pi} F^{top,B}(\acute{t}^{\alpha})
\end{equation}
Then from comparing the tree level terms we gain the closed topological string coupling $g_{top,B}$
\begin{equation}
g_{top,B}=\pm\dfrac{4\pi i}{C'X'^0}
\end{equation}
As a result of (55) and (58) one can write 
\begin{eqnarray}
ln Z_{mixed}^{IIB}=-\pi Im(F^{IIB})=F^{top,B}+\bar{F}^{top,B}
\end{eqnarray}
and hence
\begin{equation}
Z_{mixed}^{IIB}=\vert Z^{top,B} \vert^2
\end{equation}

\section{Conclusion}
\label{sec:outlook}
We predicted that due to the presence of T-duality the original OSV conjecture which was formulated for type IIA N=2 extremal black holes and type A closed topological string theory, holds in exactly the same way for type IIB N=2 exremal black holes. The only difference is that one should replace the partition function of type A closed topological string theory by that of type B closed topological string theory.
\section*{Acknowledgement}
\label{sec:ack}
We would like to thank professor H. Arfaei, A. Tavanfar, and C. Vafa for their valuable suggestions on the manuscript. This work is part of Farzad Omidi's M.S. thesis at Sharif University of Technology and he would like to thank the Department for giving him this opportunity.

\end{document}